\documentclass[journal=ancac3,manuscript=article]{achemso}

\usepackage{chemformula} 
\usepackage[T1]{fontenc} 
\usepackage{easyReview}

\author{Deepak K Sharma}
\affiliation[University]
{Laboratoire Interdisciplinaire Carnot de Bourgogne, UMR 6303 CNRS, Universit\'e de Bourgogne, 9 Avenue Alain Savary, 21000 Dijon, France}
\alsoaffiliation{Current address: Institute of Materials Research and Engineering (IMRE), Agency for Science, Technology and Research (A*STAR), 2 Fusionopolis Way, Innovis \#\ 08-03, Singapore 138634, Republic of Singapore}
\author{Adrian Agreda}
\affiliation[University]
{Laboratoire Interdisciplinaire Carnot de Bourgogne, UMR 6303 CNRS, Universit\'e de Bourgogne, 9 Avenue Alain Savary, 21000 Dijon, France}
\alsoaffiliation{Current address: ELORPrintTec, All\'ee Geoffroy Saint-Hilaire, F-33600 Pessac, France}
\author{Florian Dell’Ova}
\author{Konstantin Malchow}
\affiliation[University]
{Laboratoire Interdisciplinaire Carnot de Bourgogne, UMR 6303 CNRS, Universit\'e de Bourgogne, 9 Avenue Alain Savary, 21000 Dijon, France}
\alsoaffiliation{Current address: Laboratory of Quantum Nano Optics, EPFL Lausanne, 1015 Lausanne,
Switzerland}
\author{G\'erard Colas-des-Francs}
\author{Erik Dujardin}
\author{Alexandre Bouhelier}
\affiliation[University]
{Laboratoire Interdisciplinaire Carnot de Bourgogne, UMR 6303 CNRS, Universit\'e de Bourgogne, 9 Avenue Alain Savary, 21000 Dijon, France}
\email{alexandre.bouhelier@u-bourgogne.fr}

\title[An \textsf{achemso} demo]
  {Memristive control of plasmon-mediated nonlinear photoluminescence in Au nanowires}


\begin{document}

\begin{tocentry}

\includegraphics{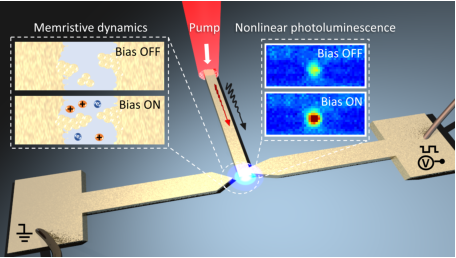}

\end{tocentry}

\begin{abstract}
Nonlinear photoluminescence (N-PL) is a broadband photon emission arising from non-equilibrium electron distribution generated at the surface of metallic nanostructures by an ultrafast pulsed laser illumination. N-PL is sensitive to surface morphology, local electromagnetic field strength, and electronic band structure making it relevant to probe optically excited nanoscale plasmonic systems. It also has been key to access the complex multiscale time dynamics ruling electron thermalization. Here, we show that the surface plasmons mediated N-PL emitted by a gold nanowire can be modified by an electrical architecture featuring a nanogap. Upon voltage activation, we observe that N-PL becomes dependent to the electrical transport dynamics and can thus be locally modulated. This finding brings an electrical leverage to externally control the photoluminescence generated from metal nanostructures, and constitutes an asset for the development of emerging nanoscale interface devices managing photons and electrons. 
\end{abstract}

\section{Introduction}
The illumination of metallic nanostructures by ultrashort laser pulses at near-IR wavelengths induces a broadband up-converted emission spanning the visible spectral regime. Under high-intensity femtosecond excitation, this nonlinear signal has been shown to be dominated by a thermal luminescence continuum emitted by out-of-equilibrium surface electrons with transiently elevated temperatures~\cite{John2015,saavedra2016,roloff2017,malchow2021,sivan2023,deAbajo:23}. Indeed, following the absorption of the light pulse, the temperature of the electron gas at the surface of the metal can reach thousands of Kelvins for about a picosecond~\cite{demichel2016,deAbajo:23} and radiates before equilibrating with the phonons~\cite{Baida11,hou2018,suemoto2021}. This nonlinear photoluminescence (N-PL) has been studied on a wide range of plasmonic geometries and materials~\cite{beversluis2003,okamoto2006,quidant2008,Xu2014,Baudrion2016,roloff2017,karki2018,grossmann2019,DellOva:23}.An important feature is the dramatic signal enhancement observed from hot-spots favored by the presence of localized resonances (e.g. at nanoparticles~\cite{bouhelier05PRL,Hecht2005}). Such local signal enhancements are observed because the luminescence emanates from hot thermal electrons generated at the surface of the metal~\cite{Boyd1986,agreda2019spatial,deAbajo:23}. Further, in structures sustaining surface plasmon propagation (e.g. nanowires, cavities), N-PL has been shown to be spatially distributed throughout the modal landscape~\cite{Viarbitskaya2015,agreda2019spatial}. Hence, N-PL has been a key observable to unlock ultrafast light-matter interactions, and this response found numerous usages. For instance, it has been used as a luminescent marker in bioimaging~\cite{wang2005,durr2007}, or as an observable to develop plasmonic logic gates\cite{kumar2021}.

The sensitivity of the N-PL to surface electrons and morphology motivates a closer examination of the signal in the presence of intricate field-driven surface effects involving atomic-scale dynamics and charge transport. These nuances serve as crucial elements for advancing the next generation of nanoscale devices in applications including energy harvesting\cite{zhang2017photochemistry}, computing and memory components\cite{yang2013memristive,lanza2022memristive}, tunable metasurfaces\cite{thyagarajan2017metasurface}, and beyond.

In this study, we show that N-PL, spatially distributed by a propagating surface plasmon in a Au nanowire (AuNW), can be locally sensitive to tiny conductance changes and atomic-scale restructuration of a nanogap. Under the application of electrostatic bias pulses across the nanogap, we observe a progressive evolution of the plasmon-mediated N-PL that eventually leads to a drastic voltage-driven modulation of its intensity. We establish that the light emission directly translates the electronic dynamics and atomic-scale movements within the nanogap.

\section{Results and discussion}

\subsection{System description}
\begin{figure}
    \centering
    \includegraphics{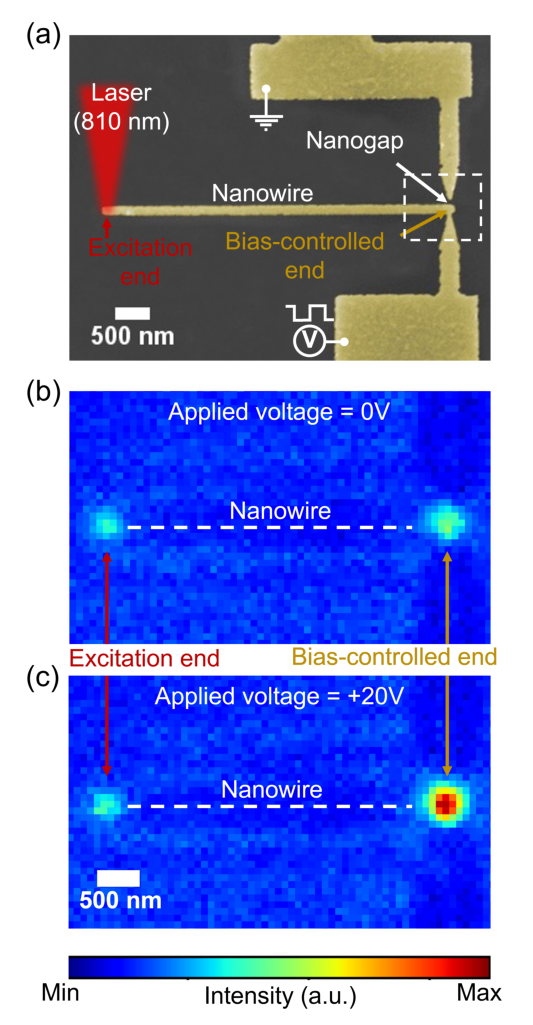}
    \caption{Overview of the AuNW system and its operation: (a) The device configuration seen from a colorized SEM image of a typical AuNW system. A tightly focused laser excitation at the left end of the AuNW is used to launch a propagating surface plasmon at the pump wavelength (810 nm). The distal end is electrostatically controlled by a tapered electrode placed at 50 nm distance from the AuNW surface. (b) Spatial distribution of the N-PL. The signal arises from the extremity directly excited by the focused laser as well as from the distal end via the mediation of a surface plasmon. Here the bias is $V =0$~V. (c) Same as in (b) for $V =20$~V. A localized enhancement of the signal at the bias-controlled end is readily seen. A dashed white line guides the long axis of the nanowire.}
\label{fig:1}
\end{figure}
Figure~\ref{fig:1}(a) is a scanning electron micrograph (SEM) of a AuNW and the in-plane electrode architecture. The electrical potential of the AuNW is applied via a first tapered electrode electrically connected to its extremity. A second tapered Au electrode acting as a ground is adjacent to it with a gap of tens of nanometers. The fabrication process involves standard electron beam lithography patterning and thin film deposition ($1~\rm{nm}$ Cr, $\sim90~\rm{nm}$ Au) on a standard glass coverslip~\cite{agreda2019spatial}. The AuNW system is then encapsulated in a polymer layer (Poly(methyl methacrylate) - PMMA, thickness $\sim 200~\rm{nm}$). The PMMA acts as a protective layer against atmospheric contaminants. The nanowires used in our experiments have cross-section of approximately $\sim 180~\rm{nm}$ in width, $\sim 90~\rm{nm}$ in thickness, and they support a bound surface plasmon mode which propagates at the excitation wavelength~\cite{song2017selective}. The surface plasmon is launched by edge scattering when the laser is focused to the left extremity of the AuNW~\cite{weeber1999plasmon}. Focusing is achieved with the help of a high numerical aperture (N.A.) objective (N.A. $= 1.49$). Figure S1 in the Supporting Information presents details and a schematic of the optical and electrical measurement setup.

Figure~\ref{fig:1}(b) is a wide-field optical image of the N-PL distribution captured by a charge-coupled device (CCD) when the AuNW is excited at its left end, and when no bias is applied ($V =0$~V). The laser is tuned at 810 nm wavelength and emits 140 fs pulse at a $80~\rm{MHz}$ repetition rate. The estimated optical peak power at the focus is $\sim 7.5~\rm{GW/m^2}$. The laser and the second harmonic generated by the Au surface are rejected from the detection using spectral filtering. A low-intensity halogen lamp illuminates the sample during the image acquisition in order to visually recognize the footprint of the system (dotted lines). The image shows a clear N-PL response from the excitation extremity where the laser intensity is focused, and a second emission region located at the right end of the AuNW, near the nanogap. Here, the nanogap is about 50~nm wide and this AuNW extremity is referred to as the bias-controlled end. The N-PL emitted at the bias-controlled end is produced by end-face scattering of the surface plasmon traveling in the nanowire at the pump laser wavelength~\cite{agreda2019spatial}.  The N-PL spectrum emitted from the bias-controlled end is presented in Figure S2 (Supporting Information). The spectrum confirms that with the laser power used here, N-PL originates from thermal hot-electrons~\cite{sivan2023,deAbajo:23}. An estimation of the temperature of the electronic distribution is inferred from a fit of the spectrum (Figure S2, supporting information) using Planck's law~\cite{agreda2020,Zhu2020,malchow2021}. We find the electron temperature at the bias-controlled end to be $T_e=1250$~K. This is a temperature somewhat cooler than previously reported values obtained from spectra measured directly under the laser focus~\cite{agreda2020,malchow2021}. This is expected because the N-PL at the bias-controlled end of the nanowire is remotely generated from the laser focus through the mediation of a lossy surface plasmon. When a positive electrostatic bias ($V =+20$~V, Figure~\ref{fig:1}(c)) is applied to the AuNW with respect to the ground electrode, a$~\sim 100\%$ enhancement of the N-PL signal is observed at the bias-controlled end. The bias modulation of the plasmon-generated N-PL is visualized, in real-time, when square electrostatic pulses(+20~V amplitude, 0.5~Hz frequency) are applied as seen in the Supplementary Movie (M1, Supporting information). 

Interestingly, the modulation of the N-PL signal is not observed on a pristine Au nanogap, but appears after a repeated electrical stress applied to the nanogap.  In the upcoming sections, we delve into this activation process and explore how a voltage-driven dynamics influences the N-PL at the bias-controlled end.

\subsection{Electrical activation of the nanogap}
\begin{figure}
    \centering
    \includegraphics[width=0.8\linewidth]{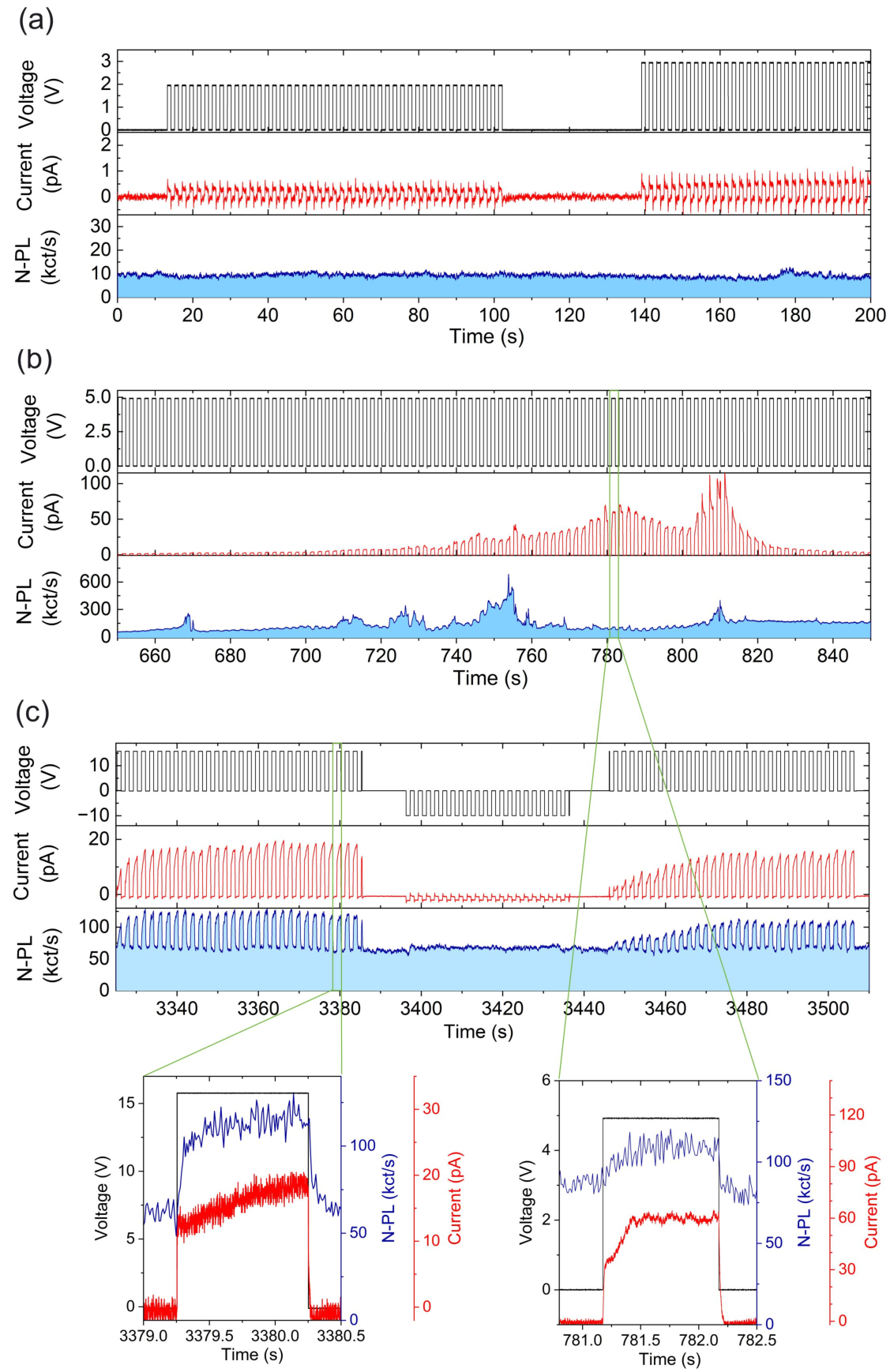}
    \caption{(a-c) Snapshots of the time series showing the evolution of modulation amplitudes in spatially filtered nonlinear photoluminescence from bias-controlled end of AuNW and electric current passing though the nanogap under electrostatic bias pulse sequences. Single-pulse zoom-in of the time traces are shown at the bottom.}
\label{fig:2}
\end{figure}

We initiate the activation process with a $V = +2$~V positive electrostatic bias pulse sequence applied to the pristine Au nanogap. We incrementally increase the voltage for subsequent pulse sequences. Throughout this activation process, we simultaneously record both the N-PL emission rate (expressed in kct/s) from the bias-controlled end (area indicated by the white dashed rectangle in Figure~\ref{fig:1}(a)), and the electric current, $I_{\rm{g}}$, flowing through the nanogap. At the beginning of the stress sequence, there is no measurable current crossing the 50 nm-wide gap. In Figure~\ref{fig:2}, we provide a series of time series illustrating the evolution of these quantities as we apply square electrostatic bias pulses with a frequency of 0.5Hz to the nanogap. The series displayed in Figure~\ref{fig:1}(a) starts with a pulse sequence at $V = +2$~V followed by one at $V = +3$~V. $I_{\rm{g}}$ remains barely above noise level ($<$ 1 pA) and is mostly dominated by artefactual capacitive effect picked up during the leading and trailing edges of the pulse. During this pulse sequence, the N-PL signal is steady and remains unaffected by the bias. Figure~\ref{fig:2}(b) is a time series acquired after repeated stress sequences done with increasing amplitudes (steps of $V = +1$~V). Here the series starts with $V = +5$~V. An electric current is now clearly measured with an amplitude fluctuating during the pulse sequence.

Remarkably, this sequence is featuring a small but correlative modulation of the N-PL at each voltage pulse (see inset). However, the baseline signal, measured at $V = 0$~V, is erratically involving with time. It has increased from $\sim 10$~kct/s at the onset of bias application at $t = 0$~s to  $\sim 150$~kct/s around $t = 850$~s under unchanged optical illumination conditions. Because N-PL is sensitive to the surface defects~\cite{agreda2021modal}, this close to 15-fold fluctuation of the baseline underscores bias-induced morphological changes occurring in the AuNW system. Despite this reshaping of the system during the sequence affecting the overall response, the small superimposed modulation of the N-PL with the bias pulses remains observable over the entire frame. 

The situation differs in the timelines represented in Figure~\ref{fig:2}(c) obtained after a prolonged application of positive pulse sequences. The N-PL baseline value is now constant over the entire time trace suggesting that transient field-driven irreversible surface alteration of the system is no longer occurring after the activation phase. In the steady regime, a $\sim$18 pA current flowing through the device is measured at each voltage pulse. This is accompanied by a clear modulation of the N-PL between 75~kct/s at $V = 0$~V to ~120 kct/s at $V = +16$~V, corresponding to a stable 60\% enhancement of the signal. In an earlier report, we have shown that N-PL intensity increases with the local electron gas temperature ($T_e$)~\cite{agreda2020}. Dwelving on the fact that $T_e$ varies as the inverse square root of the surface electron density, $N_e$, a capacitive accumulation of positive charges at the surface of the nanowire explains the signal enhancement upon application of positive bias pulses~\cite{agreda2020}. However, Figure~\ref{fig:2}(c) features an intriguing behavior when the voltage polarity of the pulses is reversed at $t = 3390$~s. Under pulses of amplitude $V = -10$~V, the current changes its polarity, but its amplitude significantly reduced, and the N-PL is no longer affected by the bias. This outcome is contrary to expectations. Indeed, keeping the capacitive scenario in mind, this polarity should lead to a periodic quenching of the signal because of the higher electron density generated at the surface of the nanowire~\cite{agreda2020}. At $t = 3445$~s, $+16$~V voltage pulses are again applied to the system. The progressive buildup of the current amplitude over the first few pulses is concomitant to a clear restoration of the N-PL modulation. These initial observations imply the existence of additional factors influencing the electrostatic bias control of the N-PL, arguably stemming from intrinsic conduction dynamics within the nanogap, which we explore in the sections below. 

\subsection{Effect of conduction dynamics on nonlinear photoluminescence modulation} 

\begin{figure}
    \centering
    \includegraphics[width=\linewidth]{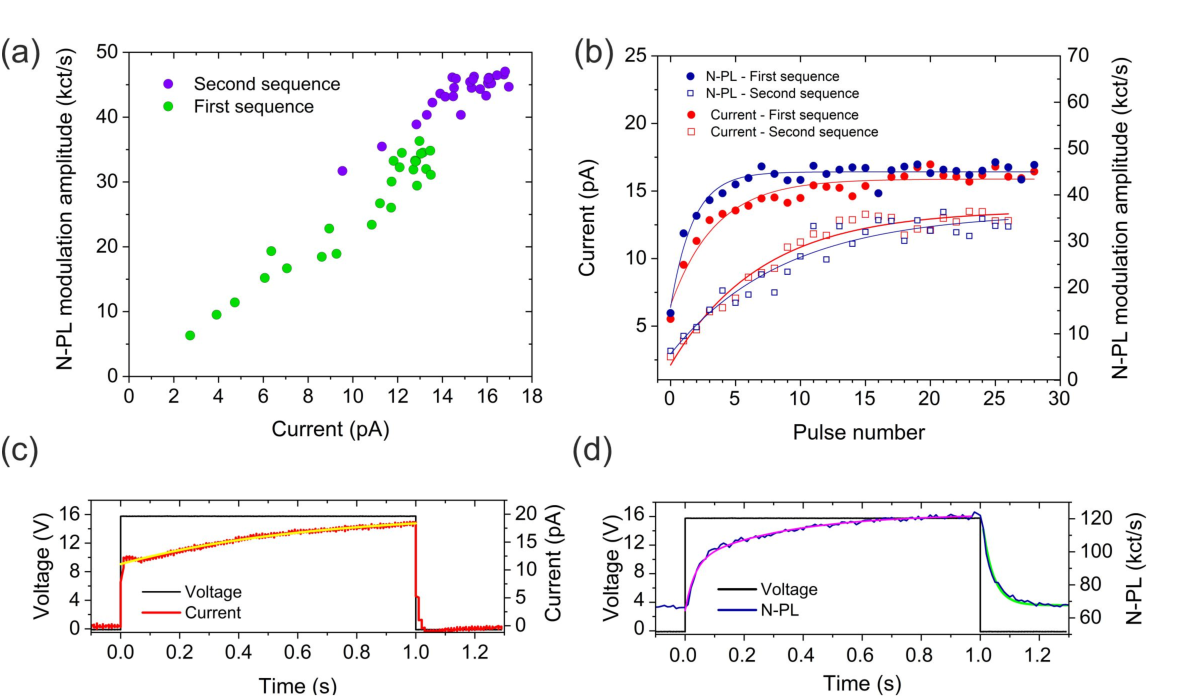}
    \caption{(a) Nonlinear photoluminescence (N-PL) modulation amplitude plotted against the electric current $I_{\rm g}$ for the two positive bias sequences shown in Figure~\ref{fig:2}(c). (b) Pulse-to-pulse evolution of the N-PL modulation amplitude and electric current amplitude for the two positive bias sequences. Solid lines are fits of the data with with as simple capacitor-like function. (c) and (d) $I_{\rm g}$ and N-PL intrapulse dynamics averaged over 29 consecutive positive bias pulses. A spurious 25 Hz component has been filtered out from teh N-PL signal. The current response is fitted with capacitor model in the slow-growth region to extract the time constant (yellow solid line). N-PL is best fitted with a double exponential after the leading bias edge (purple line). However, the decaying tail after the falling edge is well reproduced by as simple picture of a discharging capacitor (green solid line).}
\label{fig:3}
\end{figure}

The modulation amplitude (calculated as the difference between the average signal level during a bias pulse and the average baseline signal between consecutive bias pulses) of the current and the N-PL are clearly correlated. This is demonstrated in Figure~\ref{fig:3}(a) where the modulation amplitudes inferred from Figure~\ref{fig:2}(c) are plotted against each other for the two time windows corresponding to positive pulses sequences.  A linear relationship is distinctly visible indicating that the same effect rules the two responses. The underlying mechanism is also dictating the time dynamics of the signals. From Figure~\ref{fig:2}(c), it is clear that the maximum modulation amplitudes for both the current and the N-PL signals are not instantaneous but are reached after the application of a few voltage pulses. This progressive build-up of the modulation is even slower after applying the sequence of negative pulses. A comparison of the N-PL and current modulation amplitudes with the number of bias pulses is presented in Figure~\ref{fig:3}(b). We fit the data using an empirical charging function of the form  $S=a(1-\rm{exp}(-N/\tau))+c$, where $N$ is the number of pulses, and $S$ is either the current $I_{\rm g}$ or the N-PL modulation amplitude, $a$ is a weighting factor, $\tau$ is the characteristic growth coefficient and $c$ accounts for residual offset of the experiment. $a, \tau$ and $c$ are free parameters of the fit. The results (solid lines in Figure~\ref{fig:3}(b)) show that it takes between $N=3$ to $N=4$ pulses to reach maximum current modulation while 2 pulses are enough to attain the largest N-PL modulation during the first sequence of positive bias pulse. These figures are increased to 7 and 8 pulses of positive pulses (positive sequence 2) respectively after negative pulses have been applied. Such behavior unambiguously underlines the occurrence of a volatile build up of the conditions leading to a stable modulation. 

To get more insight into the intrapulse dynamics, we have averaged the current and N-PL over many pulses showing a stable modulation. Figure~\ref{fig:3}(c) reveals that after the leading bias edge, $I_{\rm g}$ suddenly sets to about 11 pA and evolves with a gradual increase over hundreds of ms. The flow of electrical charges is an unambiguous marker of the formation of conductive channels within the nanogap~\cite{Mehonic2012, Menzel19}, akin to typical memristive devices undergoing volatile resistive changes~\cite{yang2012memrisitvefilamentgrowth}. Upon electrical stress, various processes are contributing to alter in time the \ch{SiO2}/PMMA environment composing the nanogap, including defect creation produced by electron injection (e.g., oxygen vacancies), Au ions diffusion, clustering and filamentary growth~\cite{Lombardo05,Yang2008,Wei12}. Indeed, PMMA has been successfully employed as a low-cost easily-processed organic material promoting filamentary-type resistive switching~\cite{Wolf_2015,Jinesh16} whereby electrochemical reactions involving the migration and clustering metallic ions leads to the formation of an atomic conductive bridge across the gap~\cite{yang2012memrisitvefilamentgrowth,sun2019memristiveswitching}. Here, the current trace features a behavior mixing the occurrence of a non-volatile resistive state shown by the instantaneous rise of the current at the leading edge of the pulses, and an evolutionary nature materialized by a slow-growth response of hundreds of ms. The instantaneous rise of $I_{\rm g}$ preceding its growth is most probably linked to tunnel transport through the modified gap. A fit of the slow growth region using a simple model mimicking a diffusion capacitance (yellow line) yields a characteristic time of 490 ms. This time response of the current flow can be linked to the build-up of the conduction pathway in the nanogap under the influence of the applied electric field and is a strong indicator of a volatile memristive behavior~\cite{Shirakashi2019synaptic,yang2012memrisitvefilamentgrowth}. Such progressive evolution of the transport have been previously experimentally observed in wide Au nanogaps fabricated on \ch{SiO2} and were linked to the metallic diffusion and filament growth in the gap~\cite{Shirakashi2019synaptic,shirakashi2020}. We confirm the presence of metal clusters in the nanogap with scanning electron imaging of similar devices operated in air (\textit{i.e.} without the PMMA layer) and taken after extensive testing with positive and negative biases. Images are provided in Figure S3 within the Supporting Information. They clearly reveal deformed Au electrodes at the nanogap and the formation of nanometer-sized clusters, which were absent in the pre-activation SEM image. This confirms the role of field-driven surface modification on the N-PL baseline measured during the activation process the system (Figure~\ref{fig:2}(b)). 

Figure~\ref{fig:3}(d) shows the average N-PL dynamics during a pulse. Unlike the current, the enhancement of the emitted light is not instantaneous. A relatively fast raise is followed by a slower progression. A double exponential fitting (yellow curve) yields a fast exponent at 31 ms and a slower exponent at 330 ms, a latter value in line with the slow component of the current dynamics. The N-PL temporal trace shows the fast dynamics occurs also after the falling edge of the pulse. The N-PL decays to its baseline with a time constant of 36 ms, a value very close to the fast component measured at the onset of the bias pulse. Note that this decaying signal is not seen in the current trace after the bias returns to 0~V (Figure~\ref{fig:3}(c)). Within the 1~s pulse duration, the governing mechanism does not allow to reach a steady state amplitude of either of the two signals. This behavior is clear signature of the N-PL sensitivity to relaxation dynamics linked to a volatile resistive evolution and history of the nanogap. We provide in the following an interpretation how memristive dynamics may be governing the signal. Figure~\ref{fig:4} presents three sketches illustrating the activated nanogap under three biasing conditions: when the AuNW is at a positive potential (a), neutral (b), and negative potential (c) with respect to the ground tapered Au electrode.

\subsubsection{Positive bias pulse}

When a positive bias pulse is applied to the AuNW (Figure~\ref{fig:4}(a)), the cathode is partially oxidized and generates Au ions while negatively charge defects injected in the nanogap during the activation phase counter-propagate along the electric field towards their respective counter-electrode. Migration of metal atoms/ions in the switching matrix follows mutiple complex electrochemical processes and is driven by several factors such as material composition and applied field parameters~\cite{yang2014electrochemical}.  An activation of pristine nanogaps with only one polarity (Figure S4 of the Supporting Information) confirms that Au atoms/ions are detached from the positive electrode and are migrated along the electric field toward the counter electrode. For a positive bias, pre and post-activation SEM imaging reveals that a granular filament composed of nanoclusters has grown from the ground electrode and the nanowire suffered from a reshaping of its extremity. This observation is in line with previous reports of Au/\ch{SiO2}/Au systems~\cite{shirakashi2020, Shirakashi2019synaptic}.

In the present experiments, we did not apply any compliance to the electric current. Hence, complete and stable filament formation is essentially inhibited due to destructive Joule heating of the conduction path as the electric current increases. Nonetheless, we have observed complete filament growth when a 1 G$\Omega$ resistance is placed in series with the nanogap to limit the maximum current running through the device. As shown in Figure S5 of supporting information, the current trace evolves to reach as stable switching pattern characteristic of a fully non-volatile low resistive state indicative of a metallic bridge formed between the contacts. Remarkably, the modulation of the N-PL maintained during the buildup phase washes out when the low resistive state occurs. In this situation, the applied potential is dropped at the series resistance. Charged species that may be present in the gap are no longer diffusing and the electron density at the surface of the AuNW remains unaffected by the bias.  Consequently, the N-PL stays constant, as shown in Figure S5.

Coming back to the situation of a freely-evolving unprotected device, two factors likely contribute to the gradual increase of the N-PL during the pulse duration (Figure~\ref{fig:3}(c)): first, the electric field-driven volatile filamentary growth within the nanogap leads to a decrease in the effective size of the nanogap. For the same voltage applied, the electric field strength in the nanogap is thus stronger.  This contributes to lowering the surface charge density at the AuNW extremity and consequently to an enhancement of the N-PL, in agreement with our previous work~\cite{agreda2020}. The second contribution may stem from the increased density of mobile positive Au ions injected in the nanogap from the AuNW electrode during the voltage pulse. These positive metal ions near the metal surface give rise to a screened Coulomb potential and an effective lowering of the electron concentration. This may be the possible reason for the relatively slow rise of N-PL with a 31 ms time constant at the onset edge of the bias pulse compared to the current. Such tens of ms time scale are in the range of diffusion time of Ag clusters in \cite{wang2017memristors}. Once the filament gets thicker the second effect is overtaken by the capacitive effect due to a smaller gap size. 

\subsubsection{Neutral potential between positive bias pulses} 

The effect of the volatile nature of the inclusions forming the conductive pathway on the N-PL manifests itself in the decaying tail of the signal after the falling edge of the voltage pulse (Figure~\ref{fig:3}(d)). When the electric field is no longer applied, metal ions freely diffuse and relax within a sub-second dynamics and presumably form nanoclusters in the dielectric matrix~\cite{wang2017memristors,Bojun2022atomic}, as illustrated in Figure~\ref{fig:4}(b). Freely diffusing Au ions may also be reduced at negatively charged sites. It is clear that the decaying tail of the N-PL signal is ruled by such complex relaxation kinetics of Au ions.

\subsubsection{Negative bias pulse} 

\begin{figure}
    \centering
    \includegraphics[width=\linewidth]{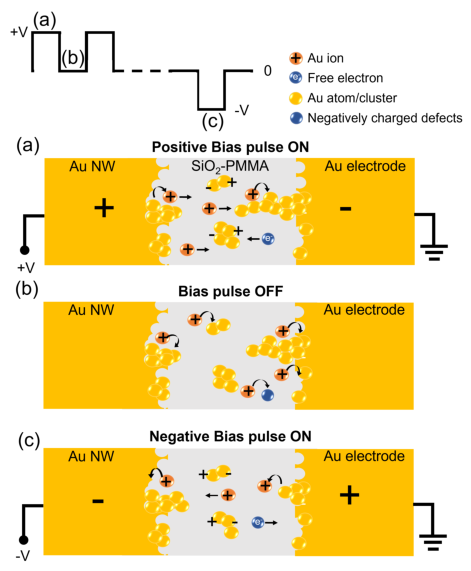}
    \caption{Schematic illustrations of positive Au atoms/ions and nanoclusters dynamics under (a) positive bias pulse, (b) between positive pulses (\textit{i.e.,} no bias), and (c) negative bias pulse.}
\label{fig:4}
\end{figure}

Upon changing the polarity of the bias, diffusing species already present in the nanogap migrate in a reverse direction and leading to a dissolution of the conductive pathways previously built-up~\cite{shirakashi2020}. This is schematically pictured in Figure~\ref{fig:4}(c). The effective widening of the nanogap results in a presumably smaller capacitive effect on the electron concentration at the surface. This is confirmed by the absence of the expected N-PL reduction~\cite{agreda2020}. The dissolution of the conduction channel is corroborated by the much lower electrical current measured for the negative sequence in Figure~\ref{fig:2}(c), where the ON conductance (defined as the ratio of mean electric current during the bias pulses with the applied voltage) is 0.13 pS compared to 0.93 pS measured during the stable response under the initial positive pulses. The reversed bias polarity acts as a partial reset process~\cite{Baker:16}.

\subsection{Investigating sub-pico Siemens conductance variation effect on N-PL modulation}
\begin{figure}
 \includegraphics[width=\linewidth]{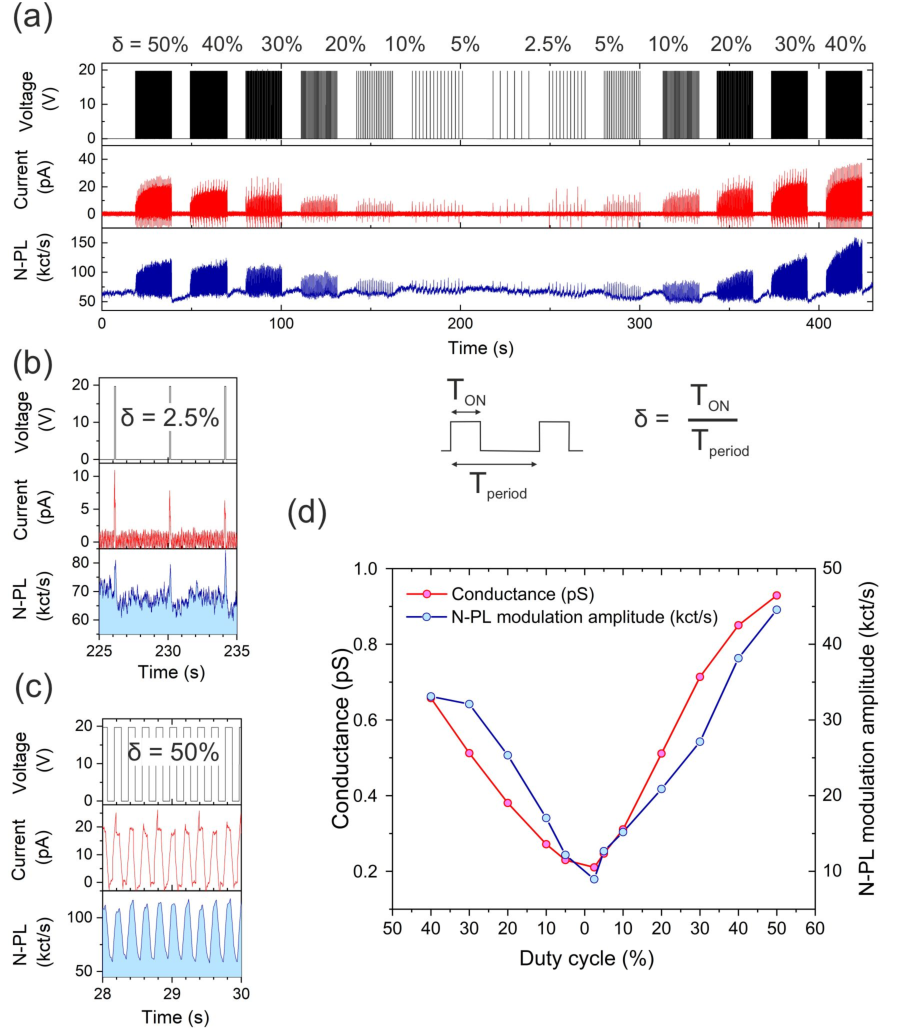}
    \caption{(a) Time series of N-PL and electric current response under 100 ms bias pulse with +20 V amplitude and varying duty cycle $\delta =T_{\rm{ON}}/T_{\rm{period}}$ shown in the inset at the top. (b) and (c) Zoomed snapshots of time series for duty cycles $\delta =2.5\%$ and $\delta =50\%$. (d) N-PL modulation amplitude and average conductance of the nanogap during the bias pulse with varying duty cycles of applied bias pulses calculated from different bias pulse sequences of time series presented in (a).}
\label{fig:5}
\end{figure}

in the following, we conduct a systematic examination of the modulation amplitude of the N-PL with varying conductance states of the nanogap under electrostatic driving. Our approach involves a control of the metallic filament growth through the application of sequences of electrostatic pulses. 
These pulses are maintained with a constant pulse width of $T_{\rm{ON}}=100$~ms and a voltage amplitude of approximately +20~V. The control parameter is here interpulse delay, \textit{i.e. }the duty cycle. Expressed as a percentage, the duty cycle $\delta$ represents the ratio of the pulse width ($T_{\rm{ON}}$, fixed as 100 ms) to the period (T$_{\rm{period}}$) of the sequence as defined on the top of Figure~\ref{fig:5}. For instance, $\delta=2.5\%$ duty cycle allows the filament to undergo 100 ms of growth followed by a relaxation time of 3.9~s, enabling self-diffusion of the species contributing to the volatile conductive state. Conversely, a $50\%$ duty cycle provides also a 100 ms for growth but also for relaxation. The space accessible for the mobile species is therefore limited before the next pulse arrives. Hence, a smaller duty cycle leads to a lower conductive state compared to a larger duty cycle.

Figure~\ref{fig:5}(a) displays the time traces of the electric current and N-PL response across subsequent varying duty cycles. Clearly, the measurements confirm that the formation kinetics of the conduction path dictates the amplitude of both responses. The current level and enhancement of the N-PL remain modest for small duty cycles (i.e., longer relaxation times), but drastically increases when T$_{\rm{ON}}$ enables the memristive dynamics to occur. The luminescence baseline at $V=0$~V remains fairly constant during this long time frame spanning over 400~s. Any drastic surface modification the bias-controlled end of the AuNW would be seen as a fluctuating signal at $V=0$~V, as in Figure~\ref{fig:2}(b). Detailed snapshots of the time series are visually presented in Figures~\ref{fig:5}(b) and (c) highlighting discernible differences in N-PL intensities and electric current for bias pulses with the extreme duty cycles. Figure~\ref{fig:5}(d) plots the N-PL modulation amplitude alongside the conductance (defined as the ratio of mean electric current during the bias pulse to the applied voltage) for varying duty cycles. Plotted values are averages across the sequences.  Here too, the conductance and the N-PL modulation amplitudes are sharing the same dependence on the diffusive dynamics of the memristive gap. These findings systematically underscore the influence of sub-pico Siemens conductance variations resulting from bias-driven filament growth and relaxation on N-PL emitted from the extremity of the nanowire. 

\subsection{Conclusion}
We presented a comprehensive experimental study of the influence of memristive dynamics on a plasmon-assisted nonlinear photoluminescence emitted by Au nanowire optically excited by focused ultrafast laser pulses. Local nonlinear photoluminescence measurements correlated with electric current flow show that field-induced dynamics leading to metallic filament formation and dissolution in the nanogap enables to modulate the intensity of the nonlinear surface response with a sub-pS sensitivity to conductance variations. These observations demonstrate that nonlinear photoluminescence can be a sensitive probe for atomic scale dynamics ruling emerging memristive technologies. We believe the system architecture explored in the present work for plasmon-assisted manupilation of electron temperature at electrically controlled remote sites will be an important platform to explore diverse applications where external controllability, agility and reconfigurability are required. They include next generation of electrically-tuned optical antennas~\cite{zurak2023direct}, and interface nanoscale devices mixing electronic and photonic functionalities~\cite{Hamdad2022}, and hot-electron management and supervision~\cite{Kang:23} .

\begin{acknowledgement}

This work has been partially funded by the French Agence Nationale de la Recherche (ANR-20-CE24-0001 DALHAI and ISITE-BFC ANR-15-IDEX-0003), the EIPHI Graduate School (ANR-17-EURE-0002). Device fabrication and characterization were performed at the technological platforms ARCEN Carnot and SMARTLIGHT with the support of the French Agence Nationale de la Recherche under program Investment for the Future (ANR-21-ESRE-0040), the Région de Bourgogne Franche-Comté, the European Regional Development Fund (FEDER-FSE Bourgogne Franche-Comté 2021/2027), the CNRS and the French RENATECH+ network.

\end{acknowledgement}

\begin{suppinfo}

The following files are available free of charge.
\begin{itemize}
  \item Movie M1: Real-time visualization of the hot-electron luminescence modulations under electrostatic bias pulses.
  \item Supporting information document: Electrical and optical characterization setup, Hot-electron luminescence spectrum, SEM images of nanogap for correlation study, Hot-electron luminescence modulation response for nanogap transitioning from high resisitve state to conductive state. 
\end{itemize}

\end{suppinfo}

\bibliography{references}

\end{document}